\documentclass[twocolumn,showpacs,amsmath,prl]{revtex4-1}

\usepackage{color}
\usepackage{graphicx}
\usepackage{epsfig,psfrag}
\usepackage{amsmath}
\usepackage{amssymb}
\usepackage{amsbsy}
\usepackage{amsthm}
\usepackage{amsfonts}
\usepackage{wasysym}
\usepackage{bbm}
\usepackage{tabularx}
\usepackage{euscript}
\usepackage{color}
\usepackage{enumerate}
\usepackage{amsfonts}
\usepackage{exscale}
\usepackage{bbold}
\usepackage{float}

\usepackage[colorlinks,citecolor=blue]{hyperref}

\renewcommand{\v}[1]{{\boldsymbol{#1}}}

\newcommand{\Eq}[1]{equation~(\ref{#1})}
\newcommand{\Fig}[1]{Fig.~\ref{#1}}

\newcommand{\<}{\langle}
\renewcommand{\>}{\rangle}

\newcommand{\Tr}{{\rm Tr}}

\newcommand{\be}{\begin{eqnarray}}
\newcommand{\ee}{\end{eqnarray}}

\newcommand{\w}{\omega}

\newcommand{\ra}{\rightarrow}
\newcommand{\s}{{\sigma}}
\def\bea{\begin{eqnarray}}
\def\eea{\end{eqnarray}}
\def\bra#1{\left\langle#1\right|}
\def\ket#1{\left|#1\right\rangle}
\def\avg#1{\left\langle#1\right\rangle}

\def\Tr{\mathrm{Tr}}

\def\Eq#1{Eq.~(\ref{#1})}
\def\Fig#1{Fig.~\ref{#1}}

\begin{document}
\title{Electronic and phononic properties of a two dimensional electron gas coupled to dipolar phonons via small-momentum-transfer scattering}
\author{Zi-Xiang Li$^{1,2}$, T. P. Devereaux$^{3,4}$}
\author{Dung-Hai Lee$^{1,2}$}\email{Corresponding author: dunghai@berkeley.edu}

\affiliation{
$^1$ Department of Physics, University of California, Berkeley, CA 94720, USA.\\
$^2$ Materials Sciences Division, Lawrence Berkeley National Laboratory, Berkeley, CA 94720, USA.\\
$^3$ Stanford Institute for Materials and Energy Sciences, SLAC National Accelerator Laboratory, Menlo Park, CA 94025, USA.\\
$^4$ Department of Materials Science and Engineering, Stanford University, Stanford, CA 94305, USA.
}

\begin{abstract}
We study the electron and phonon spectral functions as well as superconductivity for a two dimensional electron gas couped to dipolar phonons via small-momentum transfer scattering. The results reported here are obtained through sign-problem-free quantum Monte-Carlo simulation. Hence aside from modeling there is no further approximations. Our results can be relevant to the high-temperature superconductor at the
interface between a unit-cell-thick FeSe and SrTiO$_3$.
\end{abstract}

\maketitle

\noindent{{\bf Introductions}}\\
The unusual replica band observed in angle-resolved photoemission spectroscopy (ARPES)\cite{JJLee, DHLee} and its possible connection to the large pairing gap in unit-cell thick FeSe film on SrTiO$_3$ (STO)\cite{Xue,Liu,Tan,He,DLF,JFJia,YYWang,CTang-2015,Chen,Zhao,DH-2018,Hoffman-2017} has attracted much interests recently. In Ref.\cite{JJLee, DHLee} the replica band is attributed to the strong coupling between the FeSe electron and a high-frequency longitudinal optical (LO) phonon branch in the STO substrate\cite{JJLee,DHLee}. In Ref.\cite{JJLee,DHLee} it is proposed that due to such strong coupling a phonon can be simultaneously excited as the incoming photon knocks out an electron in the photoemission process. In such interpretation the electron-phonon interaction must be close to (but no need to be exactly) forward scattering in order for the replica band dispersion to follow the main band dispersion closely (an experimental fact). A rationale is provided for the origin of such near-forward  electron-phonon interaction, namely, such interaction is due to the FeSe electrons being scattered by the dynamic dipole potential exerted by the LO phonons in question. In addition, due to the known large Born effective charge of the LO phonon in question\cite{Vanderbilt,Millis}, it is argued that despite the fact that the electron-phonon interaction is limited to a narrow range of the momentum transfer, the resulting Fermi-surface-averaged electron-phonon interaction parameter $\lambda$ can be significant.

Based on the notion described above it has been proposed in Ref.\cite{JJLee,DHLee} that a cooperation exists between a pure electronic
pairing mechanism that operates in, e.g., the top surface of the K dosed FeSe films (the K dosing results in electron doping on the top surface),  and the above mentioned across-interface electron-phonon interaction. The result is an enhanced superconducting pairing gap (or gap-opening temperature) when the FeSe film is grown on an STO substrate. Note that for FeSe/STO the FeSe film is also electron doped. The doping is believed to originate from the oxygen vacancies in STO near the interface. This proposal is supported by subsequent sign-problem-free quantum Monte Carlo simulations where the electronic-pairing mechanism is assumed to be caused by spin-fluctuations\cite{Li}.

Recently several papers raised questions about the proposed explanation for the replica band. For example, in Ref.\cite{Nekrasov}  it was proposed that the replica band may simply be a renormalized FeSe band. This proposal has been ruled out by a recent isotope experiment where the energy separation between the replica and the main band was observed to undergo an expected red shift in samples where O$^{16}$ was replaced by O$^{18}$ near the FeSe-STO interface.\cite{Feng}. In another paper\cite{Jandke} it was argued that if the near forward electron-phonon scattering were strong, small-momentum phonons should suffer linewidth broadening, which is not observed by electron energy loss (EELS) experiments. A third paper\cite{Sawatsky} proposed a scenario that the replica band may be due to the energy loss of the escaped photoelectrons. 
According to this proposal the cause of the replica band seen in ARPES is exactly the same as the energy loss peak observed in EELS. Based on such notion it was proposed that ``ARPES=1/2 EELS''\cite{Sawatsky}. Moreover, using RPA approximation, Ref.\cite{Millis2} argued that in the presence of electron Coulomb interaction the proposed electron-phonon interaction  is screened, hence the enhancement of SC  should be negligible.

Since the questions raised in Ref.\cite{Jandke,Sawatsky,Millis2} are important and interesting we decide to preform an approximation-free calculation, namely sign-problem-free determinant quantum Monte Carlo (DQMC) simulations, to study the interaction between a two dimensional electron gas (which models the FeSe electrons) with a 2D array of dynamic dipoles (which models the LO phonon at the interface). However, despite our desire to do so, we can not address the question raised in Ref.\cite{Millis2} because the inclusion of the Coulomb interaction among the electrons introduces the sign problem in our DQMC simulation.

Our main results are summarized as follows. (1) Clear replica bands are observed in the electron spectral function via the inclusion of small-momentum electron-phonon interactions, qualitatively similar to that observed in experiments.\cite{JJLee} (2) Despite the substantial near-forward electron-phonon scattering the linewidth  of the small-momentum phonons is not significantly
broadened due to the fact that at zero momentum the phonon couples to a conserved quantity, namely, the total electronic charge. (3) The phonon dispersion is also not significantly affected by the electron-phonon interaction, in agreement with EELS experiments. (4) The energy separation between the replica band and the main band is blue shifted from the phonon energy, increasing near linearly with the electron-phonon interaction strength. Such a blue shift has been observed in Ref.\cite{Feng} when the ARPES result is compared with the EELS data obtained from the same sample. This blue shift is a good way to test whether the replica band seen in ARPES is due to the same mechanism that causes the energy loss peak in EELS.  (5) The electron-phonon interaction {\it by itself} does not cause a large pairing gap. However, in the presence of preexisting pairing field, it enhances the superconducting gap. If one interprets the  pairing field as due to an electronic Cooper-pairing mechanism, this  result is consistent with the notion of cooperative pairing. In the following we describe these results in more detail.\\

\noindent{\bf The model} \\
Our model consists of three parts: (1) an electron band structure, (2) a branch of high frequency phonons, (3) the dipole interaction between the electrons and phonons. Specifically, the Hamiltonian is given by:
\bea
H = H_e + H_p + H_{ep}
\label{model}
\eea
where,
\bea
H_e &=& - t \sum_{\avg{ij},\sigma} (\psi^\dagger_{i,\sigma}\psi_{j,\sigma} + h.c) - \mu \sum_i \hat{n}_{i,\sigma} \nonumber\\
H_{p} &=& \sum_{i} (\frac{\hat{P^2}}{2M} + \frac{K}{2} \hat{X}_i^2) + \frac{J}{2}(\hat{X}_i-\hat{X}_j)^2 \nonumber\\
H_{ep} &=& \alpha \sum_{i,j} \Phi(i-j) \hat{n}_i \hat{X}_j
\eea
Here, $\psi^\dagger_{i,\sigma}$ creates a spin $\s$ electron on site $i$ of a square lattice, $\mu$ is chemical potential and $\hat{n}_i$ is electron occupation number operator on site $i$. In reality the bandstructure of the FeSe film consists of two electron pockets. In the two-Fe Brillouin zone these pockets are folded on top of each other. However, in Ref.\cite{Yan} it was shown that there is no observable avoided crossing when the Fermi surfaces of these two pockets intersect. Hence the one-Fe, ``unfolded'', Brillouin zone is a good description of the bandstructure. Under this condition the small-momentum electron-phonon interaction generated by the smooth dipole potential only scatters the electrons within each Fermi pocket and thus the two electron pockets can be treated independently.

Our model bandstructure consists of only one connected Fermi surface. We choose it to be that of a nearest-neighbor tight-binding model. Consequently the center of the Fermi pocket is situated at $\v k=(0,0)$. To model the near-forward electron phonon scattering we chose the momentum range of electron-phonon interaction to be approximately 4\% of the circumference of the Fermi surface. In reality the size of Fermi pockets in FeSe/STO is only a small fraction of the Brillouin zone area. This would requires very large lattice sizes in the DQMC calculation. Since what is important is the ratio between the momentum range of the electron-phonon interaction and the circumference of the Fermi surface, we scale up the size of the Fermi pockets in order to make the calculation more feasible.

Specifically we choose the tight-binding hopping matrix element to be $t= 40$ meV. The chemical potential is chosen so that Fermi energy is $E_F = 70$ meV and the Fermi momentum is $k_F \approx \pi/2$ (in unit of inverse lattice constant). The $H_p$ in \Eq{model} is the phonon Hamiltonian. There $\hat{X}$ is the phonon displacement operator and $\hat{P}$ is its conjugate momentum. Note that the displacement of the phonon is treated as a scalar rather than a vector, where we interpret it as the dipole displacement in the direction perpendicular to the interface. At $\v q=0$ the frequency of the phonon is $\w_0 = \sqrt{K/M}$. We choose the parameters so that $\hbar\w_0 = 100$ meV and the phonon bandwidth set by $J$ is $W_p=20$ meV.

The $\Phi(i-j)$ in the third line of \Eq{model} is the dipole potential exerted on the electrons due to the phonons.  It is given by:
\bea
\Phi(i-j) = \frac{1}{((i_x - j_x)^2+(i_y-j_y)^2+q_0^{-2})^{\frac{3}{2}}}
\eea
In momentum space, the electron-phonon interaction is given by:
\bea
H_{ep} = \frac{1}{\sqrt{N}} \sum_{\v k,\v q}  g(\v q) \hat{X}_{\v q} c^\dagger_{\v k+\v q} c_{\v k}
\eea
 where $g(\v q) = 2\pi \alpha q_0 e^{-|\v q|/q_0}$ is the electron-phonon coupling constant. It  peaks at $\v q=0$ with the momentum range equal to $q_0$.  For convenience, we define the dimensionless parameter characterizing  the strength of the electron-phonon mediated electron-electron attraction as $\lambda=\left[2\sum_{\v k,\v q} |g(\v q)|^2 \delta(\epsilon_{\v k})\delta(\epsilon_{\v k-\v q})\right]/\left[N K \sum_{\v k} \delta(\epsilon_{\v k})\right]$,where $\epsilon_{\v k}$ is band dispersion in $H_e$ of  \Eq{model}. The results we shall report in the rest of the paper are produced using $q_0=0.3$ which amounts to $\sim 4\%$ of the circumference of the electron Fermi surface. The $\alpha$ in $g(\v q)$, hence the value of $\lambda$,  is an adjustable parameter.

The Holstein model with on-site electron-phonon coupling has been studied by DQMC simulations extensively \cite{Scalapino,Assaad-2007,Devereaux2, Devereaux,Johnston-2015,Kivelson,Scalettar1,Weber,Ziyang1,Scalettar2,Ziyang2}. Here we perform DQMC simulations \cite{Li,Kivelson2,Berg1,Berg2,Berg3,AV,Li2,Ziyang3,Ziyang4} to study \Eq{model}, where the electron-phonon interaction is long-ranged. Due to the long-range nature of the interaction, our simulation is considerably more time consuming than those in Ref. \cite{Scalapino,Assaad-2007,Devereaux2, Devereaux,Johnston-2015,Kivelson,Scalettar1,Weber,Ziyang1,Scalettar2,Ziyang2,Kivelson3}. Due to the time reversal symmetry ($T^2=-1$) the model in \Eq{model} is sign-problem-free\cite{Congjun,Li-18} for DQMC\cite{Sorella-1989,White-1989,Assaad-2005}. This allows us to study \Eq{model} at zero  and low temperatures on fairly big lattices without further  approximations. The largest system size we have studied is $L=14$, where $L$ is the linear dimension of the lattice. \\

\begin{figure}[t]
\includegraphics[height=2.15in]{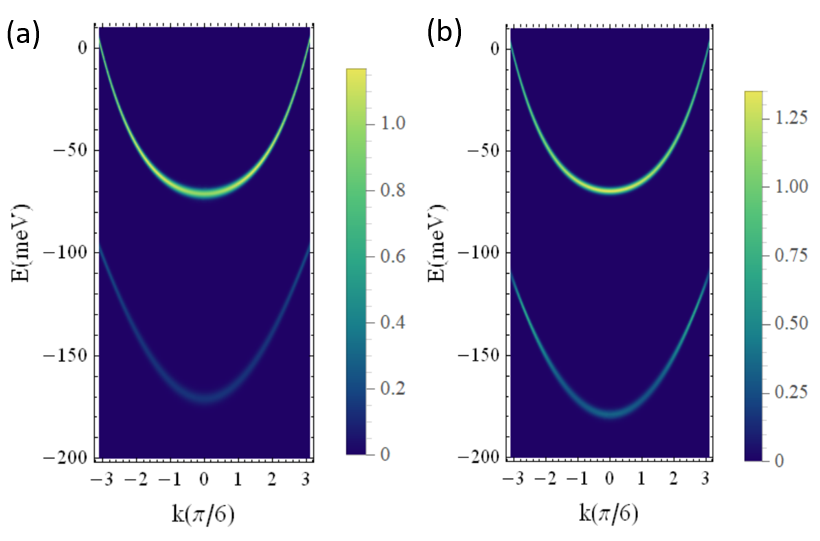}
\caption{The electron spectral function calculated for (a) $\lambda = 0.3$ and (b) $\lambda = 0.5$. The upper/lower bands are the main/replica bands, respectively. In order to obtain continuous color images we have fit the DQMC data, including the band dispersion, the linewidth, and peak intensity as a function of momentum with analytic functions. The details of the fitting are given in the supplementary materials.
}
\label{fig1}
\end{figure}

\noindent{\bf Results(1): the replica band}\\
Here we calculate the electron  spectral function at zero temperature using ``projector QMC''.  In particular we compute the imaginary-time electron Green's followed by stochastic analytical continuation\cite{Sandvik,Beach,Sandvik2} to extract the real-frequency spectral function. We leave the details of projector QMC algorithm to the supplementary materials. The system size in this  calculation is $L=12$. In \Fig{fig1}, we present the electron spectral function along a particular momentum cut, $k_y=0$, for $\lambda = 0.3$ and $\lambda = 0.5$. The results show a replica band following the dispersion of the main band but shifted toward higher binding energy. The energy splitting of the replica and the main band is approximately 100meV, the energy of the phonon at $\v q=0$.  The intensity of replica band at $\lambda = 0.5$ is slightly stronger than that at $\lambda = 0.3$. This result is qualitatively consistent with the ARPES result of Ref. \cite{JJLee}. It is also qualitatively consistent with the result of earlier perturbative calculations \cite{Johnston,Johnston-2016}. \\

\begin{figure}[t]
\includegraphics[height=1.55in]{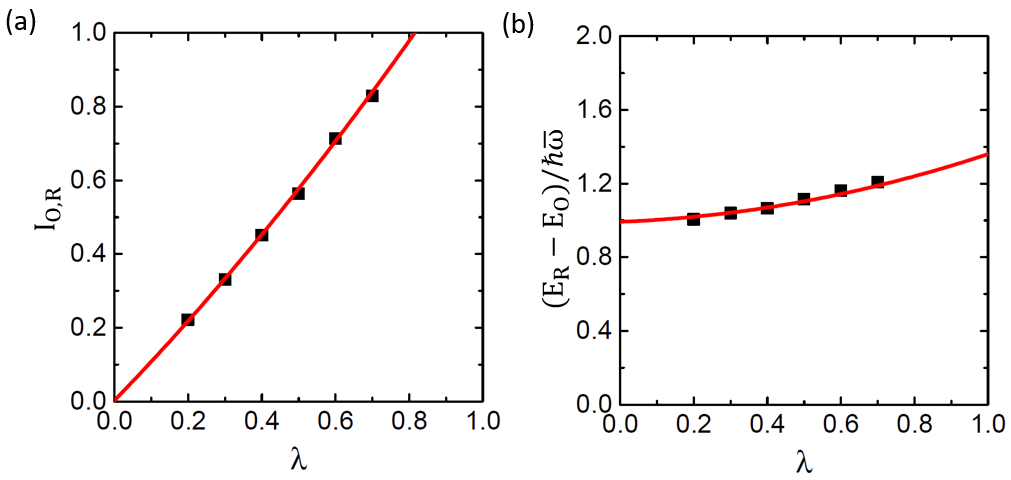}
\caption{(a) The replica to main band intensity ratio $I_{0,R}$ as a function of $\lambda$. For a given momentum the intensity is defined as the peak value of the spectral function. The red curve is a polynomial, $ a+b\lambda+c\lambda^2$, fit to the data. Here $a= 0.003 \pm 0.023 ,b= 1.031\pm 0.152 , c= 0.236 \pm 0.175 $, which indicates that linear term is dominant. (b) The ratio between the energy splitting between the replica and the main bands and the weighted average of the phonon energy as a function of $\lambda$. The red curve is a polynomial, $ a+b\lambda+c\lambda^2$, fit to the data. Here $a=0.992\pm 0.042, b= 0.141\pm 0.102, c= 0.4671\pm 0.095$.}
\label{fig2}
\end{figure}

\noindent{\bf Result(2): the band  splitting and the intensity ratio}\\
In \Fig{fig2}(a), we plot the ratio of the replica band to the main band intensity as the function of $\lambda$. The result shows an approximately linear relation, qualitatively consistent with the results from earlier perturbative calculations \cite{Johnston,Johnston-2016}. In \Fig{fig2}(b) we show the ratio between the replica-main band energy splitting and the  phonon energy as a function of $\lambda$. To account for the non-zero phonon dispersion and the momentum dependent electron-phonon coupling,  we define a weighted average phonon energy as: $\bar{\w}=\int_{BZ}d^2q \exp[-|\v q|/q_0] \w_{\v q}/\int_{BZ}d^2q \exp[-|\v q|/q_0]$,  where $\w_{\v q}$ is phonon dispersion.  The result shows the energy splitting between replica and main bands is always blue shifted relative to the phonon energy, and the shift increases with $\lambda$. We note that had we used the phonon energy at $\v q=0$ the blue shift will be even larger.  Such a blue shift is consistent with the recent experimental results where both the ARPES band splitting and the phonon energy (determined from EELS) are measured on the same sample\cite{Feng}. This  experimental result challenges the scenario proposed in \cite{Sawatsky} where the replica-main band splitting should be exactly the same as the energy of the phonon at $\v q=0$.\\

\begin{figure}[t]
\includegraphics[height=2.5in]{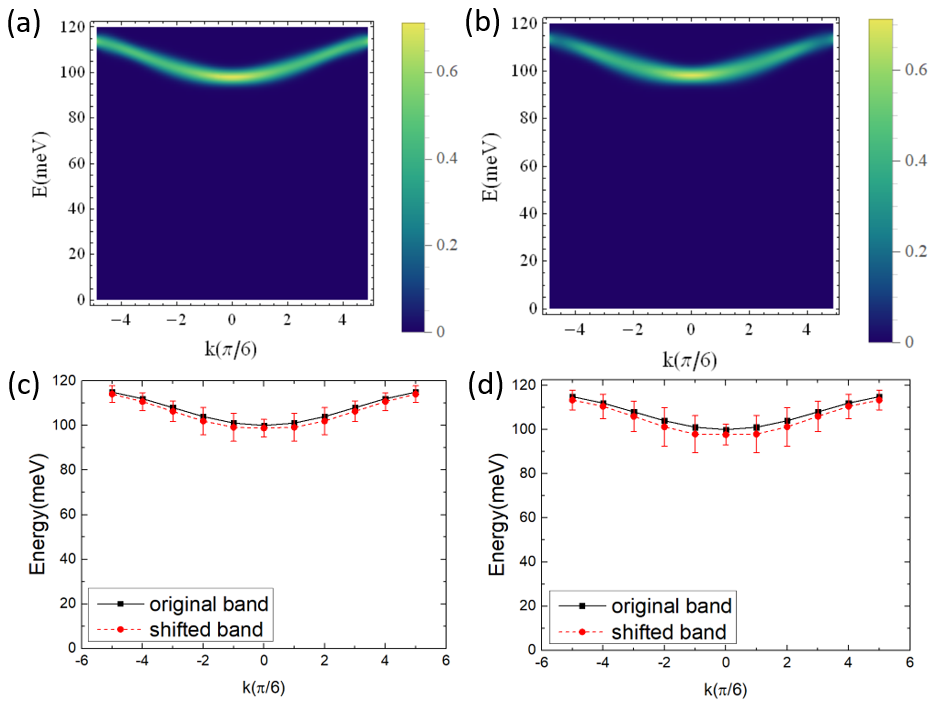}
\caption{The phonon spectral functions calculated for (a) $\lambda = 0.3$ and (b) $\lambda = 0.5$. Like Fig.1, in order to obtain continuous color images we have fit the DQMC data with analytic functions. The details of the fitting are given in the supplementary materials. Phonon linewidth and the  energy renormalization (red shift) as function of momentum for (c) $\lambda = 0.3$ and (d) $\lambda=0.5$. The black solid lines represent the bare dispersion and red dashed lines represent the renormalized dispersion. The ``error bars'' around the renormalized dispersion are the phonon linewidth. The results in this figure are obtained at a non-zero $k_BT=4 meV$.  }
\label{fig3}
\end{figure}

\noindent{\bf Result (3): the Phonon spectrum}\\
Due to the strong small-momentum electron-phonon interaction it is argued that the linewidth of small-momentum phonon should be large. However, this expectation is inconsistent with  recent electron energy loss spectroscopy \cite{Jandke,Jiandong-2016}. This fact has been used to question the validity of the attribution of the replica band to the near-forward electron-phonon interaction\cite{Jandke}.

Here we study the phonon dispersion and linewidth through sign-problem-free DQMC. In order to capture the thermal broadening of phonon linewidth we perform the calculation at non-zero temperature. Again, the phonon spectral function is obtained from the imaginary-time phonon propagator via analytical continuation. We set the temperature to be $k_BT=t/10$, which amounts to 4 meV, and the lattice size to $L=12$. The phonon dispersion and the line broadening at  $\lambda = 0.3$ and $0.5$ are shown in \Fig{fig3}(a) and \Fig{fig3}(b), respectively. Clearly the predicted substantial linewidth broadening near $\v q=0$ is absent.  In \Fig{fig3}(c) and \Fig{fig3}(d), we present raw data for phonon dispersion and linewidth (plotted as the ``error bars'') for $\lambda = 0.3$ and $0.5$. Although small, the phonon softening and linewidth both increase with increasing electron-phonon coupling strength. Importantly the phonon linewidth at zero momentum does not show substantial enhancement despite the electron-phonon coupling constant peaks at $\v q=0$. In fact, quite opposite to the expectation, the phonon linewidth and energy softening are slightly suppressed as $\v q
\ra 0$. This is due to the fact that at $\v q=0$ the phonon couples to a conserved quantity, namely the total electronic charge.  Of course,  when the phonon momentum is much larger than  $q_0$ the linewidth is also small, as expected. We believe the spurious linewidth broadening in early perturbative  calculation is due to the failure of the approximation to observe the Ward identity. Later, in Ref.\cite{Johnston,Johnston-2016}, electron Coulomb interaction is introduced to explicitly suppress the charge fluctuations. Upon doing so, the phonon linewidth at small momentum is suppressed within Eliashberg approximation.
We have also varied the temperature in the range of $k_B T=t/15 {\rm~to~} t/5$, namely 2.7-8.0 meV, to study the temperature dependence of phonon linewidth. The results are presented in the Supplementary Materials, which clearly show the phonon linewidth increases with temperature. This temperature dependence is consistent with EELS experiments (we have also calculated the phonon linewidth at zero temperature and find it is negligible). \\

\begin{figure}[t]
\includegraphics[height=1.45in]{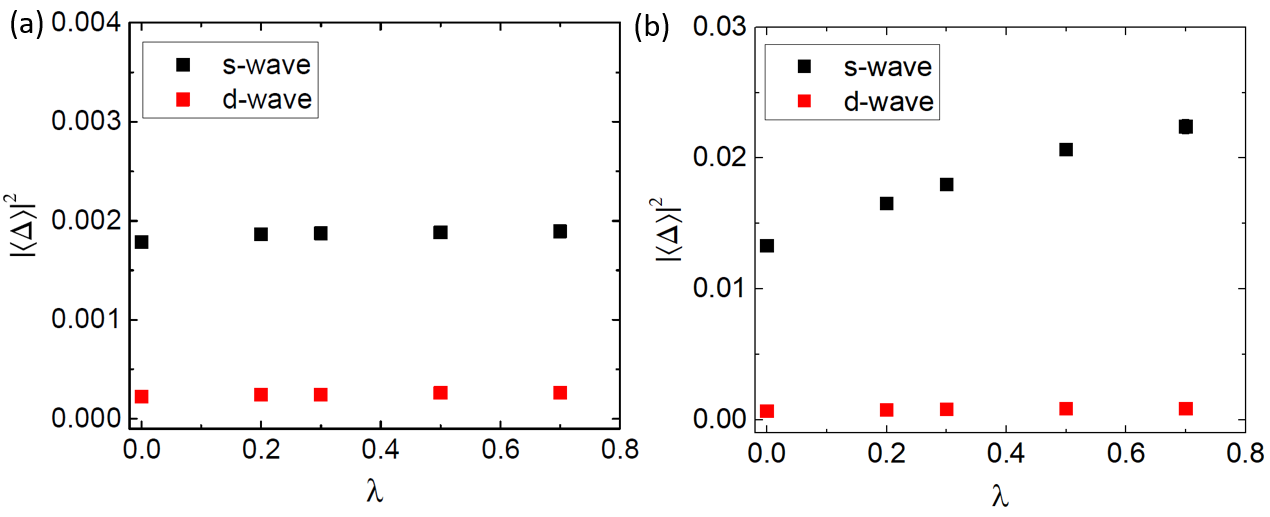}
\caption{(a) The square of the $s$-wave and $d$-wave superconducting order parameter as a function $\lambda$.  In the absence of other interactions the superconducting order parameters are small and extrapolate to zero in the thermodynamic limit (see the supplementary materials). (b) The square of the $s$ and $d$-wave superconducting order parameters as the function $\lambda$. Here a small $s$-wave pairing field $\Delta\sum_i (c_{i\uparrow}c_{i\downarrow}+h.c)$ is added to mimic the pairing induced by other electronic mechanisms. The above results are obtained in lattice with linear dimension $L=12$. }
\label{fig4}
\end{figure}

\noindent{\bf Results (4): superconductivity}\\
In the last result section we discuss the effects of small-momentum electron-phonon interaction in (a) triggering Cooper pairing\cite{Johnston-2016,Dolgov}, and (b) enhancing Cooper pairing triggered by other pairing interactions. In \Fig{fig4}(a), we plot the square of average superconducting order parameters, $|\<\Delta\>|^2$, with $s$ and $d$-wave symmetries caused by the small-momentum electron-phonon interaction {\it alone} as a function of $\lambda$ for L=12. The data shows very weak pairing, if it exists at all. We have performed finite-size scaling analysis to extract the SC order parameter in the thermodynamical limit in the supplementary materials. According to this analysis the extrapolated SC order is either zero or negative within the error bars. This implies the small-momentum electron-phonon interaction can not induce high-temperature superconductivity by itself. This is also contrary to the result of previous perturbative studies\cite{Johnston-2016,Dolgov}.

To check whether the small-momentum electron-phonon interaction can enhance Cooper pairing triggered by other electronic mechanisms, e.g., spin fluctuations\cite{JJLee,DHLee,Li}, we turn on a pair field to simulate the effect of the  pairing due to other electronic mechanisms. Due to the fact that we are dealing with a single electron pocket this pair field is chosen to have $s$-wave symmetry.
Thus we modify the  Hamiltonian in \Eq{model} to $H\ra H + \Delta \sum_i (c_{i\uparrow}c_{i\downarrow}+h.c)$  and compute the superconducting  order parameter at different values of $\lambda$. Fortunately  the revised Hamiltonian is still sign-problem-free, but it requires us to represent the fermion path integral in a Majorana representation\cite{Li-15,Li-16,Lei,Xiang,Wei}.

The resulting superconducting order parameter is shown in \Fig{fig4}(b) (here $\Delta$ is set to $0.25t $ which amounts to 10 meV). This result clearly demonstrates the ability of the small-momentum electron-phonon interaction to enhance Cooper pairing caused by other interactions. Moreover, the percentage of the order parameter enhancement is roughly consistent with (but slightly smaller than) the maximum enhancement in the gap opening temperature observed in experiments. The reason that small-momentum electron-phonon interaction by itself cannot mediate strong Cooper pairing but can enhance the Cooper pairing caused by other interaction is as follows. The small-momentum electron-phonon interaction can only induce coherence between Cooper pair in  a momentum state $(\v k,-\v k)$ with that in $(\v k+\delta\v q,-\v k-\delta \v q)$ ($|\delta\v q|$ is small). This pairing interaction does not substantially distinguish different pairing channels (e.g. $s$-wave, $d$-wave or even triplet pairing). As a result the superconducting order parameter can fluctuate  among them causing the suppression of the ordering temperature and order parameter. However if another pairing interaction has already selected a pairing channel the small-momentum electron-phonon interaction can strengthen the Cooper pairing.\\

\noindent{{\bf Concluding remarks}}\\
We have studied a two dimensional electron gas interacting with high-frequency phonons through small-momentum transfer electron-phonon interaction. Aside from modeling our calculation is free of any approximations. Our results give support to the interpretation that the observed replica band in unit-cell-thick FeSe on SrTiO$_3$ is due to the small-momentum transfer interaction between the FeSe electrons and a branch of polar  STO phonons. In addition, by studying the superconducting correlation function we conclude that while the small-momentum electron-phonon interaction can not induce strong Cooper pairing by itself, it
can enhance the Cooper pairing triggered by other pairing interactions.
\\

\noindent{{\bf Acknowledgement}} \\

This work was primarily funded by the U.S. Department of Energy, Office of Science, Office of Basic Energy Sciences, Materials Sciences and Engineering Division under Contract No. DE-AC02-05-CH11231 (Quantum Material Program KC2202).  The computational part of this research is supported by the U.S. Department of Energy, Office of Science, Office of Advanced Scientific Computing Research, Scientific Discovery through Advanced Computing (SciDAC) program. Z.X.L. and D.H.L. also acknowledge support from the Gordon and Betty Moore Foundation's EPIC initiative, Grant GBMF4545.

\begin{widetext}
\section{Supplementary Information}

\renewcommand{\theequation}{S\arabic{equation}}
\setcounter{equation}{0}
\renewcommand{\thefigure}{S\arabic{figure}}
\setcounter{figure}{0}
\renewcommand{\thetable}{S\arabic{table}}
\setcounter{table}{0}

\subsection{I. How are \Fig{fig1} and \Fig{fig3}(a,b) generated ?}
Since our calculation is done on finite lattices we can only compute the spectral functions of electrons and phonon at discrete lattice momenta {\bf delete listing k points since it looks like the simulation is only for 1D} $k=(\frac{2\pi n}{L},0)$. In order to obtain continuous color images, we use analytical functions to fit the band dispersion and band broadening obtained from the DQMC simulations.
From the calculated spectral function, we extract three sets of numbers:
$E(\v k),I(\v k),\Gamma(\v k)$. Here $E(\v k)$ is the energy at which $A(k,E)$ reaches the maximum value for a given $\v k$. $I(\v k)$ is the peak value of spectral function, and $\Gamma(\v k)$ is the width of spectral peak. For a given momentum cut polynomial functions of $k$ are used to fit these three quantities. We find that six-order polynomial function is sufficient to generate a good fit for the electronic $E(k),I(k),\Gamma(k)$  and eight-order polynomial function is sufficient to fit the  those for the phonon. Using the fitted continuous functions for $E(k)$,$I(k)$ and $\Gamma(k)$ we reconsruct the continuous electron spectral function as: \be A(k,E)= \frac{I(k)}{\sqrt{2\pi\Gamma(k)}} e^{\frac{-(E-E(k))^2}{2\Gamma(k)}},
\ee where gaussian distribution of spectrum function around the peak is assumed. Such obtained $A(k,E)$ are plotted in \Fig{fig1} and \Fig{fig3}(a,b).

\subsection{II. Projector Quantum Monte Carlo and finite-temperature Quantum Monte Carlo}
The projector quantum Monte Carlo\cite{Sorella-1989,White-1989,Assaad-2005} is a DQMC algorithm. It is used to study the ground state properties. In projector QMC, the ground state expectation value of an observable  can be evaluated as:
\bea
\avg{\hat{O}} = \frac{ \bra{\psi_{0}} O \ket{\psi_0}}{ \avg{\psi_{0} \mid \psi_{0} } }   = \lim_{\theta\rightarrow \infty} \frac{ \bra{\psi_T}e^{-\theta H } O e^{-\theta H} \ket{\psi_T}}{ \bra{\psi_T} e^{-2\theta H} \ket{\psi_T}}
\eea
where $| \psi_0\> $ is the true ground state and $ |\psi_T\>$ is a trial state which has a non-zero overlap with ground state.  In our calculation $|\psi_T\>$ is taken as the electronic ground state in the absence of electron-phonon interaction. The parameter $\theta$ is set to $30/t$, and we have checked that the various observables are unchanged within the statistical uncertainty upon further increment of $\theta$.
In addition, standard Trotter decomposition is employed to discretize $\theta$ into many small intervals $\Delta\theta$.  In our calculation $\Delta\theta$ is set to $0.1/t$ for small $\lambda$ and $0.05/t$ for big $\lambda$. We have and also checked that results do not change upon further decreasing $\Delta\theta$.

For non-zero temperature calculations we evaluate the following average of observables:
\bea
\avg{\hat{O}} = \frac{\Tr[\hat{O}e^{-\beta \hat{H}}]}{\Tr[e^{-\beta \hat{H}}]}
\eea
where $\beta$ is inverse temperature. Like described in the last paragraph $\beta$  is also divided into small intervals $\Delta \tau$ via the Trotter decomposition. In our calculation $\Delta\tau$ is set to $0.1/t$ for small $\lambda$ and $0.05/t$ for big $\lambda$. Convergence against further decrease of $\Delta\tau$ has been checked.

In both the zero and finite temperature calculations we employ both single-site and global updates to ensure the statistical independence in our Monte-Carlo sampling. Specifically in the global update, we update the phonon fields at all the imaginary time associated with a given site. We perform $2-5$ global updates in every sweep of DQMC.

\begin{figure}
\includegraphics[height=2.5in]{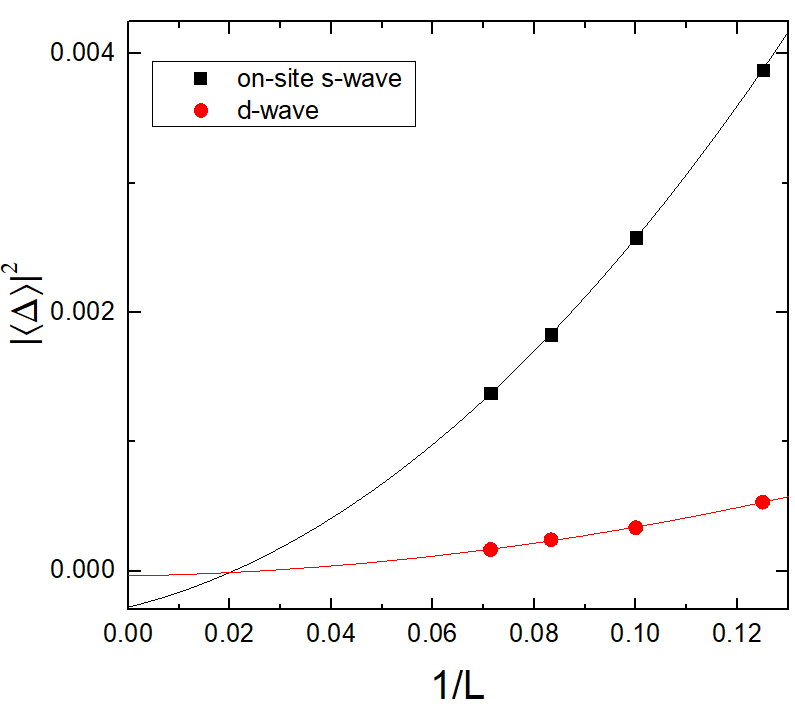}
\caption{Finite-size scaling analysis of on-site $s$-wave and $d$-wave SC structure factors for $L=8,10,12,14$. The square of the average superconducting order parameters are fitted by second-order polynomial functions. In both cases the structure factors are extrapolated to zero within error bars, which suggests that the induced pairing by forward-scattering electron-phonon coupling is negligible.}
\label{figSM1}
\end{figure}

\subsection{III. Finite-size scaling analysis of the square of superconducting order parameter induced by the small-momentum electron-phonon interaction alone}

To extract the superconducting  order induced solely by the small-momentum electron-phonon interaction in thermodynamics limit, we perform finite-scaling analysis of the square of the order parameter associated with $s$ and $d$ wave pairing for $L=8,10,12,14$ with $\lambda$ fixed at $0.5$. The results are fitted by second-order polynomial functions of $1/L$, the extrapolation  of which is the result in  the thermodynamic limit. The  fittings are shown in \Fig{figSM1}. Clearly the superconducting order is negligible in thermodynamic limit.\\

When a pair field is added, the computation is constrained to much smaller system sizes due to the enlarged matrix size in the Majorana representation. It is thus difficult to obtain reliable finite-size scaling results. However, the significant enhancement of the order parameter in finite systems provide strong evidence that small-momentum electron-phonon interaction can indeed enhance Cooper pairing induced by other mechanisms.

\subsection{IV. Temperature dependence of phonon linewidth}
We have varied the temperature to study the temperature dependence of the phonon linewidth. The temperature range in our simulation is $k_B T = t/5$ to $t/15$, namely $2.7-8.0$ meV. We have fixed $\lambda=0.5$ and have evaluated the width at the momentum where phonon linewidth is largest. The results clearly show that phonon linewidth increases with temperature, as plotted in \Fig{figSM2}. The trend of increasing phonon linewidth with temperature is consistent with EELS experiments and earlier perturbative calculations.

\begin{figure}
\includegraphics[height=2.5in]{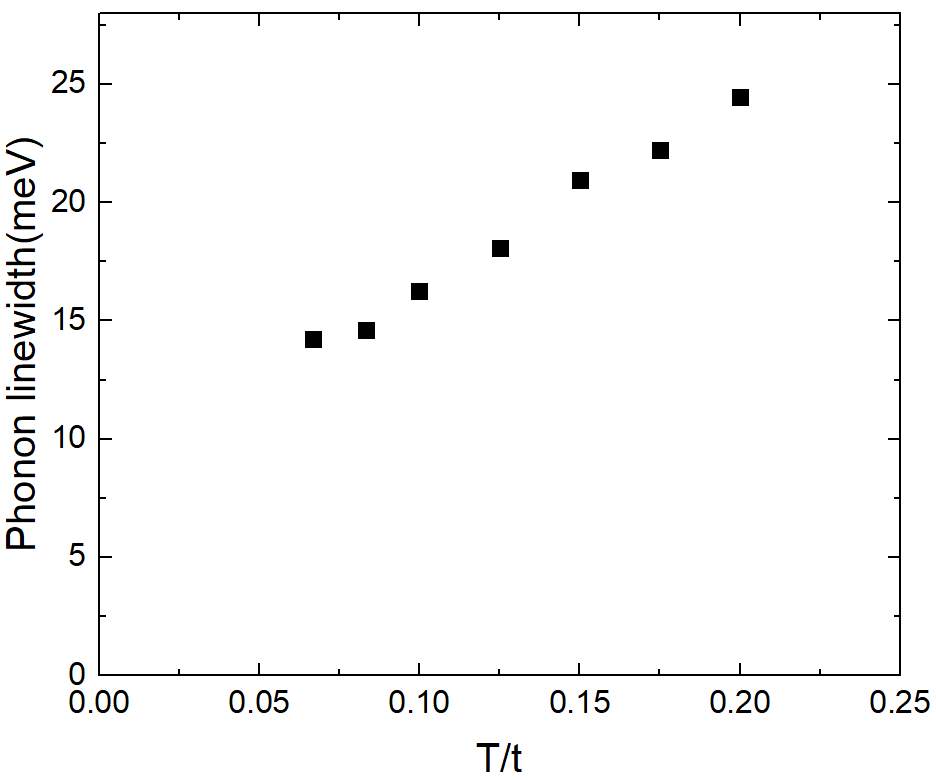}
\caption{The phonon linewidth as the function of temperature for $\lambda=0.5$. The phone linewidth increases roughly linearly with temperature. }
\label{figSM2}
\end{figure}

\end{widetext}

\end{document}